\documentclass[aps,showpacs,nofootinbib,superscriptaddress]{revtex4}

\usepackage{amsmath}
\usepackage{graphicx}
\usepackage{float}
\usepackage{bm}
\usepackage{url}

\begin{document}

\title{Sivers Asymmetry in the pion induced Drell-Yan process at COMPASS within TMD factorization}
\author{Xiaoyu Wang}
\affiliation{School of Physics, Southeast University, Nanjing 211189, China}
\author{Zhun Lu}
\email{zhunlu@seu.edu.cn}
\affiliation{School of Physics, Southeast University, Nanjing 211189, China}

\begin{abstract}

We investigate the Sivers asymmetry in the pion-induced single polarized Drell-Yan process in the theoretical framework of the transverse momentum dependent factorization up to next-to-leading logarithmic order of QCD. Within the TMD evolution formalism of parton distribution functions, the recently extracted nonperturbative Sudakov form factor for the pion distribution functions as well as the one for the Sivers function of the proton are applied to numerically estimate the Sivers asymmetry in the $\pi^- p$ Drell-Yan at the kinematics of the COMPASS at CERN. In the low $b$ region, the Sivers function in $b$-space can be expressed as the convolution of the perturbatively calculable hard coefficients and the corresponding collinear correlation function, of which the Qiu-Sterman function is the most relevant one. The effect of the energy-scale dependence of the Qiu-Sterman function to the asymmetry is also studied. We find that our prediction on the  Sivers asymmetries as functions of $x_p$, $x_\pi$, $x_F$ and $q_\perp$ is consistent with the recent COMPASS measurement.
\end{abstract}

\maketitle

\section{Introduction}

The Sivers function~\cite{Sivers:1989cc} is a transverse momentum dependent~(TMD) parton distribution function~(PDF), which describes the asymmetric distribution of unpolarized quarks inside a transversely polarized nucleon through the correlation between the quark transverse momentum and the nucleon transverse spin. Because of its time-reversal-odd (T-odd) property, the Sivers function plays an important role in the understanding of the transverse spin structure of the nucleon~\cite{Barone:2010zz} within the twist-2 approximation of QCD parton model. It can also give rise to the single-spin asymmetry in various high energy scattering processes.
During the last decade, the Sivers asymmetry in semi-inclusive deep inelastic scattering (SIDIS) has been measured by the HERMES~\cite{Airapetian:2004tw,Airapetian:2009ae}, COMPASS~\cite{Alekseev:2008aa,Alekseev:2010rw,Adolph:2012sp,Adolph:2016dvl}, and Jlab Hall A~\cite{Qian:2011py} Collaborations.
The data from these experiments were utilized by
several groups~\cite{anselmino05b,efr05,cegmms,vy05,Anselmino:2008sga,Anselmino:2012aa}
to extract the quark Sivers functions of the proton.
However, the TMD framework of QCD predicts that the T-odd PDFs present generalized universality, i.e., the sign of the Sivers function measured in Drell-Yan process should be opposite to its sign measured in SIDIS~\cite{Collins:2002kn,Brodsky:2002rv,Brodsky:2002cx}. The verification of this sign change~\cite{Anselmino:2009st,Kang:2009bp,Peng:2014hta,Echevarria:2014xaa,Huang:2015vpy,Anselmino:2016uie} is one of the most fundamental tests of our understanding of the QCD dynamics and the factorization scheme, and it is also the main pursue of the existing and future Drell-Yan facilities~\cite{Aghasyan:2017jop,Gautheron:2010wva,Fermilab1,Fermilab2,ANDY,Adamczyk:2015gyk}.

Very recently, the COMPASS Collaboration has reported the first measurement of the Sivers asymmetry in the pion-induced Drell-Yan process, in which a $\pi^-$ beam was scattered off the transversely polarized NH$_3$ target~\cite{Aghasyan:2017jop}. The polarized Drell-Yan data from COMPASS, together with the previous measurement of the Sivers effect in the $W$- and $Z$-boson production from $p^\uparrow p$ collision at relativistic heavy ion collider (RHIC)~\cite{Adamczyk:2015gyk}, provide the first evidence of the sign change of the Sivers function.
The COMPASS experiment has the unique advantage to explore the sign change of the Sivers function since it has almost the same setup~\cite{Aghasyan:2017jop,Adolph:2016dvl} for SIDIS and Drell-Yan process, which may reduce the uncertainties in the extraction of the Sivers function from the two kinds of measurements.
An important issue in the comparison of observables between the SIDIS and Drell-Yan-type processes is that the typical energy scales for existing SIDIS facilities are quite different from those for the existing and planned hadron-hadron collision facilities.
To obtain reliable theoretical estimate of the Sivers asymmetry, the evolution effects must be included.
Since most of the data accumulated by the COMPASS Collaboration are at low transverse momentum of the dilepton pair, a natural choice for the analysis is TMD factorization, which is valid in the region where $q_\perp$ is much smaller than the hard scale $Q$.

The main purpose of this work is to apply the TMD factorization to present a detailed phenomenological analysis of the Sivers asymmetry in the pion induced Drell-Yan process. Particularly, we take into account the TMD evolution for both the pion distribution functions and the proton distribution functions. In the TMD formalism~\cite{Collins:1981uk,Collins:1984kg,Collins:2011zzd,Ji:2004xq}, the differential cross section can be separated to the hard scattering factors and the well-defined TMD PDFs or fragmentation functions (FFs). At some fixed energy scale, one may express the TMD PDFs or FFs as convolutions of their collinear counterparts and the perturbatively calculable $C$-coefficients. Specifically, the Sivers function in the coordinate space
(conjugate to the transverse momentum space through Fourier transformation) in the perturbative region can be written as the convolution of the $C$-coefficients and the corresponding collinear correlation functions, among which Qiu-Sterman matrix element is the most relevant one as the Qiu-Sterman function $T_{q,F}(x,x)$ appears in the leading order (in the $\alpha_s$ expansion of QCD) contribution for the structure function $\widetilde{W}_{UT}^\alpha(Q,b)$~\cite{Kang:2011mr}, and consequently it may provide the main contribution to the Single spin asymmetry. Other twist-3 correlation functions appear in the next-to-leading order corrections for the structure function $\widetilde{W}_{UT}^\alpha$~\cite{Kang:2011mr} and are ignored in our study. Therefore, we will consider the Qiu-Sterman function as the source of the corresponding collinear correlation function of Sivers function. After solving the evolution equations, the evolution from one energy scale to another energy scale is realized by an exponential factor, the so-called Sudakov form factor~\cite{Collins:1984kg,Collins:2011zzd,Collins:1999dz}, which can be separated into a perturbatively calculable part $S_\textrm{P}$ and a nonperturbative part $S_\textrm{NP}$.
In this study, we take all the $C$-coefficients and the perturbative Sudakov form factors up to the next-to-leading logarithmic~(NLL) accuracy to get reliable results.
The nonperturbative Sudakov form factors can not be calculated directly and are usually parameterized from experimental data. In Ref.~\cite{Wang:2017zym}, the nonperturbative Sudakov form factor corresponding to the pion distribution functions was extracted by using the unpolarized $\pi^- N$ Drell-Yan data from the E615 experiment~\cite{Conway:1989fs,Stirling:1993gc} at FermiLab. Here we apply the same parameterized result to estimate the Sivers asymmetry in the pion induced Drell-Yan at COMPASS.
On the other hand, the Sivers effects in SIDIS, $pp$ Drell-Yan and $W/Z$-production have been studied extensively in Refs.~\cite{Aybat:2011ta,Anselmino:2012aa,Sun:2013hua,Echevarria:2014xaa}, in which different parametrizations for the nonperturbative Sudakov form factor corresponding to the Sivers function of the proton were proposed.
In this work we adopt the expression of $S_\textrm{NP}$ from Ref.~\cite{Echevarria:2014xaa}.
For consistency we also take the parametrization from Ref.~\cite{Echevarria:2014xaa} for the collinear counterpart of the Sivers function, the so-called Qiu-Sterman function.

The rest of the paper is organized as follows. In Sec.~\ref{Sec.formalism}, we present the theoretical framework for the Sivers asymmetry in the pion induced transversely polarized Drell-Yan process within the TMD factorization. In Sec.~\ref{Sec.numerical}, we numerically calculate the Sivers asymmetry for the underlying process at the kinematics of COMPASS Collaboration using the framework set up in Sec.~\ref{Sec.formalism}. We summarize the paper in Sec.~\ref{Sec.conclusion}.

\section{Formalism of the Sivers asymmetry in Drell-Yan}

\label{Sec.formalism}

In this section, following the procedure in Ref.~\cite{Collins:2011zzd}, we review the necessary setup of the TMD factorization to obtain the theoretical expression of the Sivers asymmetry in the pion-induced Drell-Yan process:
\begin{align}
\pi^-(P_\pi)+p(P_p,S_p)\longrightarrow \gamma^*(q)+X \longrightarrow l^+(\ell)+l^-(\ell')+X,
\end{align}
where $P_\pi$, $P_p$ and $q$ represent the momenta of the $\pi^-$ meson, the proton and the virtual photon, respectively. $S_p$ is the four-vector of the target polarization. In contrast to the SIDIS process, $q$ is a time-like vector in Drell-Yan process, namely, $Q^2=q^2>0$, with $Q^2$ the invariant mass square of the lepton pair. We adopt the following kinematical variables~\cite{Collins:1984kg,Gautheron:2010wva} to express the experimental observables
\begin{align}
&s=(P_{\pi}+P_p)^2,\quad x_\pi=\frac{Q^2}{2P_\pi\cdot q},\quad x_p=\frac{Q^2}{2P_p\cdot q},\nonumber\\
&x_F=2q_L/s=x_\pi-x_p,\quad\tau=Q^2/s=x_\pi x_p,\quad y=\frac{1}{2}\mathrm{ln}\frac{q^+}{q^-}=\frac{1}{2}\mathrm{ln}\frac{x_\pi}{x_p},
\end{align}
where $s$ is the total center-of-mass~(c.m.) energy squared; $x_\pi$ and $x_p$ are the Bjorken variables of the pion and nucleon, respectively; $q_L$ is the longitudinal momentum of the virtual photon in the c.m. frame of the incident hadrons; $x_F$ is the Feynman $x$ variable; and $y$ is the rapidity of the lepton pair. Thus, $x_\pi$ and $x_p$ can be expressed as functions of $x_F$, $\tau$ and of $y$, $\tau$
\begin{align}
x_{\pi/p}=\frac{\pm x_F+\sqrt{x_F^2+4\tau}}{2},\quad x_{\pi/p}=\sqrt{\tau} e^{\pm y}.
\end{align}

The transverse single spin asymmetry for the unpolarized $\pi^-$ scattering off the transversely polarized proton Drell-Yan process can be defined as~\cite{Echevarria:2014xaa}
\begin{align}
A_{UT}=\frac{d^4\Delta\sigma}{dQ^2dyd^2\bm{q}_{\perp}}\bigg{/}{\frac{d^4\sigma}{dQ^2dyd^2\bm{q}_{\perp}}},
\label{eq:asy_Sivers}
\end{align}
where $\frac{d^4\sigma}{dQ^2dyd^2\bm{q}_{\perp}}$ and $\frac{d^4\Delta\sigma}{dQ^2dyd^2\bm{q}_{\perp}}$ stand for the spin-averaged and spin-dependent differential cross section, respectively, and $\bm{q}_{\perp}$ is the transverse momentum of the dilepton.

In general, it is convenient to solve the TMD evolution effects in the $\bm{b}$-space which is conjugate to $\bm{q}_\perp$. Thus, the structure functions in Drell-Yan process are usually expressed in the $\bm{b}$-space as products of hard scattering factor and distribution functions in the $\bm{b}$-space. The physical observables can be obtained through a Fourier transform from the $\bm{b}$-space to the $\bm{q}_\perp$.

The spin-averaged differential cross section can be written as~\cite{Collins:1984kg}
\begin{equation}
\label{eq:dsigma_UU}
\frac{d^4\sigma}{dQ^2dyd^2\bm{q}_{\perp}}=\sigma_0\int \frac{d^2b}{(2\pi)^2}e^{i\vec{\bm{q}}_{\perp}\cdot \vec{\bm{b}}}\widetilde{W}_{UU}(Q;b)+Y_{UU}(Q,q_{\perp}),
\end{equation}
where $\sigma_0=\frac{4\pi\alpha_{em}^2}{3N_CsQ^2}$ is the cross section at the tree level, $\widetilde{W}_{UU}(Q,b)$ is the spin-independent structure function in the $b$-space which contains all-order resummation results and dominates in the low $q_{\perp}$ region ($q_{\perp}\ll Q$); while the $Y_{UU}$ term provides necessary correction at $q_{\perp}\sim Q$. Hereafter, we will use the terms with a tilde to denote the quantities in $b$-space.

The spin-dependent differential cross section has the form~\cite{Kang:2011mr,Echevarria:2014xaa,Sun:2013hua}
\begin{equation}
\label{eq:dsigma_UT}
\frac{d^4\Delta\sigma}{dQ^2dyd^2\bm{q}_{\perp}}=\sigma_0\epsilon_\perp^{\alpha\beta} S^\alpha_\perp\int \frac{d^2b}{(2\pi)^2}e^{i\vec{\bm{q}}_{\perp}\cdot \vec{\bm{b}}}\widetilde{W}^\beta_{UT}(Q;b)+Y^\beta_{UT}(Q,q_{\perp}).
\end{equation}
Similarly, $\widetilde{W}_{UT}(Q,b)$ is the spin-dependent structure function in $b$-space and dominates at $q_{\perp}\ll Q$, while $Y^\beta_{UT}$ provides correction for the single polarized process at $q_{\perp}\sim Q$. The antisymmetric tensor $\epsilon^{\alpha\beta}_\perp$ is defined as $\epsilon^{\alpha\beta\mu\nu}P_\pi^\mu P_p^\nu / P_\pi\cdot P_p$, and $S_\perp$ is the transverse polarization vector of the proton target.
As we always focus on the region $q_\perp\ll Q$, we will neglect the $Y$-terms and will only consider the $W$-terms on the r.h.s of Eqs.~(\ref{eq:dsigma_UU}) and (\ref{eq:dsigma_UT}).

According to the TMD factorization~\cite{Collins:1984kg,Collins:2011zzd}, the structure functions can be expressed as the product of the well defined TMD PDFs and the process/scheme-dependent hard factors. Therefore, the structure functions $\widetilde{W}_{UU}(Q,b)$ and $\widetilde{W}_{UT}(Q,b)$ can be written as
\begin{align}
&\widetilde{W}_{UU}(Q;b)=H_{UU}(Q;\mu) \sum_{q,\bar{q}}e_q^2\tilde{f}_{1\, \bar{q}/\pi}(x_\pi,b;\mu,\zeta_F)
\tilde{f}_{1\,q/p}(x_p,b;\mu,\zeta_F),\label{eq:struc_uu}\\
&\widetilde{W}^{\alpha}_{UT}(Q;b)=H_{UT}(Q;\mu) \sum_{q,\bar{q}}e_q^2\tilde{f}_{1\, \bar{q}/\pi}(x_\pi,b;\mu,\zeta_F)
\tilde{f}_{1T\,q/p}^{\perp\alpha(\mathrm{DY})}(x_p,b;\mu,\zeta_F).\label{eq:struc_funcs}
\end{align}
Here, $\tilde{f}_{1q/H}$ is the unpolarized distribution function in the $b$-space with the soft factor subtracted in the definition of the TMD distribution functions. $\tilde{f}_{1T\,q/p}^{\perp\alpha}(x_p,b;\mu,\zeta_F)$ is the subtracted Sivers function for proton in the $b$-space, which is defined as~\cite{Echevarria:2014xaa}
\begin{align}
\tilde{f}_{1T\,q/p}^{\perp\alpha(\mathrm{DY})}(x,b;\mu,\zeta_F)=\int d^2\bm{k}_\perp e^{-i\vec{\bm{k}}_\perp\cdot\vec{\bm{b}}}\frac{k^\alpha_\perp}{M_p}
f^{\perp(\mathrm{DY})}_{1T,q/p}(x,\bm{k}_\perp;\mu),
\end{align}
where the superscript DY denotes that the quark Sivers function is the one in the Drell-Yan process, and it satisfies the relation $f^{\perp(\mathrm{DY})}_{1T,q/p}=-f^{\perp(\mathrm{DIS})}_{1T,q/p}$.

In Eqs.~(\ref{eq:struc_uu}) and (\ref{eq:struc_funcs}), $H_{UU}(Q;\mu)$ and $H_{UT}(Q;\mu)$ are the factors associated with the corresponding hard scattering, $\mu$ is the renormalization scale in the case of the collinear PDFs, and $\zeta_F$ is the energy scale serving as a cutoff to regularize the light-cone singularity of the TMD distributions. We note that, in the above definition the soft factors have been absorbed into the definition of the TMD PDFs, and the way to subtract the soft factor in the distribution function depends on the regulating scheme for the light-cone singularity\cite{Collins:1981uk,Collins:2011zzd}. In literature, two different schemes are usually applied: the Collins-11 scheme~\cite{Collins:2011zzd} and the Ji-Ma-Yuan scheme~\cite{Ji:2004wu,Ji:2004xq}, which yield scheme-dependent hard factors $H_{UU}(Q;\mu)$ and $H_{UT}(Q;\mu)$. However, after combining with TMD PDFs, the final results of the physical observables should be scheme independent.

\subsection{The spin-averaged differential cross section}

The general expression for the unpolarized structure function $\widetilde{W}_{UU}$ in terms of the unpolarized TMD PDF $\tilde{f}_{1q/H}$ for the pion and proton in the $b$-space is given in Eq.~(\ref{eq:struc_uu}). For the evolution effect of the TMD PDFs, there are two energy dependencies that should be solved, one is the $\zeta_F$-dependence and the other is the $\mu$-dependence. The former dependence is encoded in the Collins-Soper~(CS)~\cite{Collins:2011zzd} equation as
\begin{align}
\frac{\partial\ \mathrm{ln} \tilde{f}_1(x,b;\mu,\zeta_F)}{\partial\ \sqrt{\zeta_F}}=\tilde{K}(b;\mu),
\label{eq:CS}
\end{align}
while the latter one is derived from the renormalization group equation as:
\begin{align}
\frac{d\ \tilde{K}}{d\ \mathrm{ln}\mu}&=-\gamma_K(\alpha_s(\mu)),\\
\frac{d\ \mathrm{ln} \tilde{f}_1(x,b;\mu,\zeta_F)}
{d\ \mathrm{ln}\mu}&=\gamma_F(\alpha_s(\mu);\frac{\zeta^2_F}{\mu^2}),
\label{eq:RGs}
\end{align}
with $\tilde{K}$ the CS evolution kernel, and $\gamma_K$ and $\gamma_F$ the anomalous dimensions. The solutions of these evolution equations have been studied in details in Ref.~\cite{Collins:2011zzd}. Here, we only discuss the final result. The overall solution structure for $\tilde{f}_1(x,b;\mu,\zeta_F)$ is the same as that for the Sudakov form factor. Namely, the energy evolution of TMDs from a initial energy $\mu$ to another energy $Q$ is encoded in the Sudakov-like form factor $S$ by the exponential form $\mathrm{exp}(-S)$
\begin{equation}
\tilde{f}(x,b,Q)=\mathcal{F}\times e^{-S}\times \tilde{f}(x,b,\mu),  \label{eq:f}
\end{equation}
where $\mathcal{F}$ is the hard factor depending on the scheme one chooses, while the solution structure is independent on the scheme. Hereafter, we set $\mu=\sqrt{\zeta_F}=Q$ and express $f(x,b;\mu=Q,\zeta_F=Q^2)$ as $f(x,b;Q)$ for simplicity.

Since the transverse momentum dependence of the experimental observable can be determined by the $b$-dependence of the structure function through Fourier transformation, it is quite important to understand the $b$-dependence of the TMD functions.
In the large $b$ region, the dependence is nonperturbative because the operators are separated by a large distance and should contain the nonperturbative functions that can be extracted from the experimental data. While in the small $b$ region, the $b$-dependence of the TMDs is perturbatively calculable and can be expressed in terms of the corresponding collinear distribution functions.
A matching procedure must be introduced to combine the perturbative calculation at small $b$ with the nonperturbative fits at large $b$. With a parameter $b_{\mathrm{max}}$ serving as the boundary between perturbative and nonperturbative region,
a $b$-dependent function $b_\ast$ is defined to have the property $b_\ast\approx b$ at low values of $b$ and $b_{\ast}\approx b_{\mathrm{max}}$ at large $b$ values.
The typical value of $b_{\mathrm{max}}$ is chosen around $1\ \mathrm{GeV}^{-1}$ such that $b_{\ast}$ is always in the perturbative region.
There are several different $b_\ast$ prescriptions in literature~\cite{Collins:2016hqq,Bacchetta:2017gcc}. In this work we adopt the original prescription introduced in Ref.~\cite{Collins:1984kg} as $b_{\ast}=b/\sqrt{1+b^2/b^2_{\mathrm{max}}}$.

With the introduction of the $b_\ast$ prescription, the Sudakov-like form factor $S$ in Eq.~(\ref{eq:f}) can be separated into the perturbatively calculable part and the nonperturbative part
\begin{equation}
\label{eq:S}
S=S_{\mathrm{P}}+S_{\mathrm{NP}}.
\end{equation}
The perturbative part of $S$ has the form
\begin{equation}
\label{eq:Spert}
S_{\mathrm{P}}(Q,b)=\int^{Q^2}_{\mu_b^2}\frac{d\bar{\mu}^2}{\bar{\mu}^2}\left[A(\alpha_s(\bar{\mu}))
\mathrm{ln}\frac{Q^2}{\bar{\mu}^2}+B(\alpha_s(\bar{\mu}))\right].
\end{equation}
The coefficients $A$ and $B$ in Eq.(\ref{eq:Spert}) can be expanded as a $\alpha_s/{\pi}$ series:
\begin{align}
A=\sum_{n=1}^{\infty}A^{(n)}(\frac{\alpha_s}{\pi})^n,\\
B=\sum_{n=1}^{\infty}B^{(n)}(\frac{\alpha_s}{\pi})^n.
\end{align}
In this work, we take $A^{(n)}$ up to $A^{(2)}$ and $B^{(n)}$ up to $B^{(1)}$ within the NLL accuracy~\cite{Collins:1984kg,Landry:2002ix,Qiu:2000ga,Kang:2011mr,Aybat:2011zv,Echevarria:2012pw}:
\begin{align}
A^{(1)}&=C_F,\\
A^{(2)}&=\frac{C_F}{2}\left[C_A\left(\frac{67}{18}-\frac{\pi^2}{6}\right)-\frac{10}{9}T_Rn_f\right],\\
B^{(1)}&=-\frac{3}{2}C_F.
\end{align}

For the nonperturbative form factor $S_{\mathrm{NP}}$ associated with the unpolarized TMD PDF of the proton, a parameterization has been proposed in Ref.~\cite{Su:2014wpa} to study the unpolarized $pp$ Drell-Yan process:
\begin{align}
S_{\mathrm{NP}}=g_1b^2+g_2\mathrm{ln}\frac{b}{b_{\ast}}\mathrm{ln}\frac{Q}{Q_0}
+g_3b^2\left((x_0/x_1)^{\lambda}+(x_0/x_2)^\lambda\right).
\label{eq:SNP_DY_NN}
\end{align}
At the initial scale $Q^2_0=2.4\ \mathrm{GeV}^2$ with $b_{\mathrm{max}}=1.5\ \mathrm{GeV}^{-1}$, $x_0=0.01$ and $\lambda=0.2$, the parameters in Eq.~(\ref{eq:SNP_DY_NN}) are fitted to the experimental data to get the values $g_1=0.212,\ g_2=0.84, \ g_3 = 0$.
Since the nonperturbative form factors $S_{\mathrm{NP}}$ for quarks and antiquarks satisfy the following relation~\cite{Prokudin:2015ysa}
\begin{align}
S^q_{\mathrm{NP}}(Q,b)+S^{\bar{q}}_{\mathrm{NP}}(Q,b)=S_{\mathrm{NP}}(Q,b),
\end{align}
and both the initial hadrons in the collision process are nucleons,
the $S_{\mathrm{NP}}$ associated with the TMD distribution function of the initial protons can be expressed as
\begin{align}
\label{eq:SNPproton}
&S^{f_{1\,q/p}}_{\mathrm{NP}}(Q,b)=\frac{g_1}{2}b^2+\frac{g_2}{2}\ln\frac{b}{b_{\ast}}\ln\frac{Q}{Q_0}.
\end{align}

In Ref.~\cite{Wang:2017zym}, we fit the nonperturbative Sudakov-like form factor $S_{\mathrm{NP}}$ for the pion distribution function from the $\pi^- N$ Drell-Yan data~\cite{Conway:1989fs,Stirling:1993gc} with the following form
\begin{align}
S^{f_{1q/\pi}}_{\mathrm{NP}}=g^\pi_1\,b^2+g^\pi_2\mathrm{ln}\frac{b}{b_{\ast}}\mathrm{ln}\frac{Q}{Q_0}.
\label{eq:SNP_pion}
\end{align}
The parameters $g_1^\pi$ and $g_2^\pi$ are fitted at the initial energy scale $Q^2_0=2.4\ \mathrm{GeV}^2$ with $b_{\mathrm{max}}=1.5\ \mathrm{GeV}^{-1}$ as $g^\pi_1=0.082$ and $g^\pi_2=0.394$.
The perturbative form factors $S_{\mathrm{P}}$ for quarks and antiquarks have the relation of~\cite{Prokudin:2015ysa}
\begin{align}
S^q_{\mathrm{P}}(Q,b_\ast)=S^{\bar{q}}_{\mathrm{P}}(Q,b_\ast)=S_{\mathrm{P}}(Q,b_\ast)/2.
\end{align}

When $b$ is in the perturbative region $1/Q \ll b \ll 1/ \Lambda$, the TMD distribution function at fixed energy in $b$-space can be expressed as the convolution of perturbatively calculable hard coefficients and the corresponding collinear PDFs ~\cite{Collins:1981uk,Bacchetta:2013pqa}
\begin{equation}
\tilde{f}_{1q/H}(x,b;\mu)=\sum_i C_{q\leftarrow i}\otimes f_1^{i/H}(x,\mu)
\label{eq:f_fixed_engy},
\end{equation}
where $\otimes$ stands for the convolution in the momentum fraction $x$
\begin{equation}
 C_{q\leftarrow i}\otimes f_1^{i/H}(x,\mu)\equiv \int_{x}^1\frac{d\xi}{\xi} C_{q\leftarrow i}(x/\xi,b;\mu,\zeta_F)f_1^{i/H}(\xi,\mu),
 \label{eq:otimes}
\end{equation}
and $f_1^{i/H}(\xi,\mu)$ is the collinear PDF of $i$ flavor in hadron $H$ at the energy scale $\mu$, which could be a dynamic scale related to $b_\ast$ by $\mu_b=c_0/b_\ast$, with $c_0=2e^{-\gamma_E}$ and $\gamma_E\approx0.577$ the Euler Constant~\cite{Collins:1981uk}. In addition, the sum $\sum_i$ runs over parton flavors.
Here, the $b_\ast$ prescription prevents $\alpha_s(\mu_b)$ from hitting the so-called Landau pole at large $b$ regime.
In particular, the $C$-coefficients in Eq.~(\ref{eq:f_fixed_engy}) are universal among different schemes and initial hadrons.
With all the ingredients presented above, we can rewrite the unpolarized distribution function $f_1$ in the $b$-space as a function of $x$, $b$, and $Q$,
\begin{align}
\label{eq:f1}
\tilde{f}_{1q/p}(x,b;Q)&=e^{-\frac{1}{2}S_{\mathrm{P}}(Q,b_\ast)
-S^{f_{1q/p}}_{\mathrm{NP}}(Q,b)}\mathcal{F}(\alpha_s(Q))
\sum_iC_{q\leftarrow i}\otimes f_{1}^{i/p}(x,\mu_b),\\
\tilde{f}_{1q/\pi}(x,b;Q)&=e^{-\frac{1}{2}S_{\mathrm{P}}(Q,b_\ast)-S^{f_{1q/\pi}}_{\mathrm{NP}}(Q,b)}
\mathcal{F}(\alpha_s(Q))\sum_iC_{q\leftarrow i}\otimes f_{1}^{i/\pi}(x,\mu_b).
\end{align}

Substituting Eq.~(\ref{eq:f1}) into Eq.~(\ref{eq:struc_funcs}), we express the spin-averaged structure function as
\begin{align}
\widetilde{W}_{UU}(Q;b)=&H_{UU}(Q;\mu)\sum_{q,i,j} e_q^2\mathcal{F}(\alpha_s(Q))C_{q\leftarrow i}\otimes f_{1i/p}(x_p,\mu_b)\mathcal{F}(\alpha_s(Q))\nonumber\\
\times&C_{\bar{q}\leftarrow j}\otimes f_{1j/\pi}(x_\pi,\mu_b)
e^{-\left(S^{f_{1q/p}}_{\mathrm{NP}}+S^{f_{1q/\pi}}_{\mathrm{NP}}+S_\mathrm{P}\right)},
\label{eq:struc_unp}
\end{align}
where $S^{f_{1q/p}}_{\mathrm{NP}}$, $S^{f_{1q/\pi}}_{\mathrm{NP}}$ and $S_\mathrm{P}$ are given in Eqs.~(\ref{eq:SNPproton}),~(\ref{eq:SNP_pion}) and~(\ref{eq:Spert}), respectively.
Therefore, the spin-averaged differential cross section can be cast into
\begin{align}
\frac{d^4\sigma}{dQ^2dyd^2\bm{q}_{\perp}}=\frac{\sigma_0}{2\pi}\int^\infty_0dbbJ_0(q_\perp b)\widetilde{W}_{UU}(Q;b),
\label{eq:dcs_unp}
\end{align}

We note that the factors $H_{UU}(Q;\mu)$ and $\mathcal{F}(\alpha_s(Q))$ relating to the hard scattering are scheme-dependent. If we absorb them to the definition of the $C$-coefficients, the cross section in Eq.~(\ref{eq:dcs_unp}) can be arranged as
\begin{align}
\frac{d^4\sigma}{dQ^2dyd^2\bm{q}_{\perp}}=\frac{\sigma_0}{2\pi}\int^\infty_0dbbJ_0(q_\perp b)
\sum_{q,i,j} e_q^2C^\prime_{q\leftarrow i}\otimes f_{1i/p}(x_p,\mu_b)C^\prime_{\bar{q}\leftarrow j}\otimes f_{1j/\pi}(x_\pi,\mu_b)
e^{-\left(S^{f_{1q/p}}_{\mathrm{NP}}+S^{f_{1q/\pi}}_{\mathrm{NP}}+S_\mathrm{P}\right)},
\label{eq:dcs_unp_id}
\end{align}
with the new $C$-coefficients having the form~\cite{Catani:2000vq}
\begin{align}
&C^\prime_{q\leftarrow q^{\prime}}(x,b;\mu_b)=\delta_{qq^{\prime}}\left[\delta(1-x)+\frac{\alpha_s}{\pi}
\left(\frac{C_F}{2}(1-x)+\frac{C_F}{4}(\pi^2-8)\delta(1-x)\right)\right],\label{eq:cfactor_ab1}\\
&C^\prime_{q\leftarrow g}(x,b;\mu_b)=\frac{\alpha_s}{\pi}T_Rx(1-x).
\label{eq:cfactor}
\end{align}

\subsection{The spin-dependent differential cross section}

In this subsection, we present the theoretical framework of the spin-dependent differential cross section in the $\pi N$ Drell-Yan contributed by the Sivers function, following the procedure in Ref.~\cite{Echevarria:2014xaa}.
The general expression of the structure function $\widetilde{W}_{UT}$ is given in Eq.~(\ref{eq:struc_funcs}). The evolution functions for the Sivers function in the $b$-space $\tilde{f}_{1T\,q/p}^{\perp\alpha(\mathrm{DY})}$ follow the same ones in Eqs.~(\ref{eq:CS}),~(\ref{eq:RGs}) and the solution structure can also be written in the same form as that in Eq.~(\ref{eq:f}). The Sudakov form factor in the perturbative region for the Sivers function is exactly the same as the one for the unpolarized distribution function\cite{Echevarria:2014xaa,Kang:2011mr,Aybat:2011ge,Echevarria:2012pw,Echevarria:2014rua}.

In Ref.~\cite{Echevarria:2014xaa}, the authors proposed a nonperturbative Sudakov form factor in the evolution formalism, which can lead to a good description of the transverse momentum distribution for different processes such as SIDIS, DY dilepton and W/Z boson production in $pp$ collisions.
The nonperturbative Sudakov form factor $S_{\mathrm{NP}}$ in Ref.~\cite{Echevarria:2014xaa} for the Sivers function has the form
\begin{align}
S^{\mathrm{Siv}}_{\mathrm{NP}}=\left(g^{\mathrm{Siv}}_1+g^{\mathrm{Siv}}_2\mathrm{ln}\frac{Q}{Q_0}\right)b^2,
\label{eq:SNP_Sivers}
\end{align}
where the parameter $g^{\mathrm{Siv}}_1$ is relevant to the averaged intrinsic transverse momenta squared $g^{\mathrm{Siv}}_1=\langle k^2_{s\perp}\rangle_{Q_0}/4=0.071\mathrm{GeV}^2$, and $g^{\mathrm{Siv}}_2$ being $\frac{1}{2} g_2=0.08 \mathrm{GeV}^2$.

Similar to what has been done to the unpolarized distribution function in Eq.~(\ref{eq:f_fixed_engy}), in the low $b$ region, the Sivers function $\tilde{f}_{1T\,q/p}^{\perp\alpha(\mathrm{DY})}$ can be also expressed as the convolution of perturbatively calculable hard coefficients and the corresponding collinear correlation functions as~\cite{Sun:2013hua,Kang:2011mr}
\begin{equation}
\tilde{f}_{1T\,q/p}^{\perp\alpha(\mathrm{DY})}(x,b;\mu)=(\frac{-ib^\alpha}{2})\sum_i \Delta C^T_{q\leftarrow i}\otimes f^{(3)}_{i/p}(x',x'';\mu).
\label{eq:Siv_fixed_engy}
\end{equation}
Here, $\Delta C^T_{q\leftarrow i}$ stands for the hard coefficients, and
$f^{(3)}_{i/p}(x',x'')$ denotes the twist-three quark-gluon-quark or trigluon correlation
functions, among which the transverse spin-dependent Qiu-Sterman matrix element $T_{q,F}(x',x'')$~\cite{Qiu:1991pp,Qiu:1991wg,Qiu:1998ia} is the most relevant one.
As shown in Eq.~(6) in Ref.~\cite{Kang:2011mr}, in the small $b$ region the Qiu-Sterman function $T_{q,F}(x,x)$ (corresponding to the case $x^\prime = x''$) appears in the leading order (in the $\alpha_s$ expansion) contribution in Eq.~(\ref{eq:Siv_fixed_engy}):
\begin{align}
\tilde{f}_{1T\,q/p}^{\perp\alpha(\mathrm{DY})}(x,b)\big{|}_{\textrm{LO}} = \left(\frac{-ib^\alpha}{2}\right)T_{q,F}(x,x).
\end{align}
Thus $T_{q,F}(x,x)$ may provide the main contribution to the single spin asymmetry.
Other twist-3 correlation functions appear in the next-to-leading order corrections
for the structure function $\widetilde{W}_{UT}^\alpha$~\cite{Kang:2011mr} and are ignored in our study.
The relation between the Qiu-Sterman function $T_{q,F}(x,x)$ and the quark Sivers function is given by~\cite{Sun:2013hua,Kang:2011mr}
\begin{align}
T_{q,F}(x,x)=\int d^2k_\perp \frac{|k_\perp^2|}{M_p}f^{\perp{\textrm{DY}}}_{1T\,q/p}(x,k_\perp) = 2M_p\,f_{1T\,q/p}^{\perp(1) \textrm{DY}}(x), \label{eq:qs_moment}
\end{align}
which is proportional to the first transverse moment of the Sivers function $f_{1T\,q/p}^{\perp(1)}(x)$.
Therefore, in the following we use the relation in Eq.~(\ref{eq:qs_moment}) to model the $x$-dependence of
the Sivers function in terms of the phenomenological information available on the Qiu-Sterman function.
Similar to what has been done in the last subsection, the scheme-dependent hard factors are absorbed into the $C$-coefficients definition, leading to~\cite{Kang:2011mr,Sun:2013hua}
\begin{align}
\Delta C^T_{q\leftarrow q^{\prime}}(x,b;\mu_b)=\delta_{qq^{\prime}}\left[\delta(1-x)+\frac{\alpha_s}{\pi}
\left(-\frac{1}{4N_c}(1-x)+\frac{C_F}{4}(\pi^2-8)\delta(1-x)\right)\right],
\label{eq:cfactor_Siv}
\end{align}
where $C_{\bar q\leftarrow j}$ is given in Eq.~(\ref{eq:cfactor_ab1}). Now we can express the Sivers function in $b$-space as
\begin{align}
&\tilde{f}_{1T,q/p}^{\perp}(x,b;Q)=\frac{b^2}{2\pi}\sum_i \Delta C^T_{q\leftarrow
i}\otimes T_{i,F}(x,x;\mu_b) e^{-S^{\mathrm{siv}}_{\mathrm{NP}}-
\frac{1}{2}S_{\mathrm{P}}},\label{eq:siv_b}
\end{align}
and in the transverse momentum space as
\begin{align}
&\frac{k_\perp}{M_p}f_{1T,q/p}^{\perp}(x,k_\perp;Q)=\int_0^\infty db
\frac{b^2}{2\pi}J_1(k_\perp b)\sum_i \Delta C^T_{q\leftarrow i}\otimes
f^{\perp(1)}_{1T,i/p}(x,\mu_b) e^{-S^{\mathrm{siv}}_{\mathrm{NP}}-
\frac{1}{2}S_{\mathrm{P}}}.\label{eq:QS_Q}
\end{align}

In the c.m. frame of colliding hadrons, we adopt a convenient coordinate system to choose the unpolarized $\pi^-$ beam to move along the $+z$ direction, the transverse polarized proton along the $-z$ direction, the spin vector $S_\perp$ along the $y$ direction. This is consistent with the choice of the COMPASS experiments~\cite{Aghasyan:2017jop}.
We can rewrite the spin-dependent differential cross section in Eq.~(\ref{eq:dsigma_UT}) as
\begin{align}
\frac{d^4\Delta\sigma}{dQ^2dyd^2\bm{q}_{\perp}}&=\sigma_0\epsilon^{\alpha\beta} S^\alpha_\perp\int \frac{d^2b}{(2\pi)^2}e^{i\vec{\bm{q}}_{\perp}\cdot \vec{\bm{b}}}\widetilde{W}^\beta_{UT}(Q;b)\nonumber\\
&=\sigma_0\epsilon_\perp^{\alpha\beta} S^\alpha_\perp\int \frac{d^2b}{(2\pi)^2}(\frac{-ib^\beta}{2})\sum_{q,i,j}
e_q^2\Delta C^T_{q\leftarrow i}T_{i,F}(x_p,x_p,\mu_b)\nonumber\\
&\times C_{\bar q\leftarrow j}\otimes f_{1,j/\pi}(x_\pi,\mu_b)
e^{-\left(S^{\mathrm{Siv}}_{\mathrm{NP}}+S^{f_{1q/\pi}}_{\mathrm{NP}}+S_\mathrm{P}\right)}\nonumber\\
&=\frac{\sigma_0}{4\pi}\int^\infty_0dbb^2J_1(q_\perp b)\sum_{q,i,j}
e_q^2\Delta C^T_{q\leftarrow i}T_{i,F}(x_p,x_p;\mu_b)C_{\bar q\leftarrow j}\otimes f_{1,j/\pi}(x_\pi,\mu_b)
e^{-\left(S^{\mathrm{Siv}}_{\mathrm{NP}}+S^{f_{1q/\pi}}_{\mathrm{NP}}+S_\mathrm{P}\right)}.
\label{eq:dcs_Sivers}
\end{align}

\section{Numerical calculation}

\label{Sec.numerical}

In this section, we present the numerical calculation of the Sivers asymmetry $A_{UT}^{\textrm{Siv}}$ in $\pi^- p^\uparrow \to \mu^+ \mu^- +X$ at the kinematics of COMPASS Collaboration using the framework introduced above.

In order to obtain the numerical estimate of the denominator of the asymmetry given in Eq.~(\ref{eq:dcs_unp_id}), we adopt the NLO set of the CT10 parametrization~\cite{Lai:2010vv}~(central PDF set) for the unpolarized distribution function $f_1(x)$ of the proton. For the unpolarized PDF of the pion meson, we use the NLO SMRS parametrization~\cite{Sutton:1991ay}.
To estimate the numerator of the asymmetry in Eq.~(\ref{eq:dsigma_UT}), we adopt a recent parameterization~\cite{Echevarria:2014xaa} for the Qiu-Sterman functions $T_{q,F}(x,x;\mu)$ extracted from the Sivers asymmetry in SIDIS:
\begin{align}
T_{q,F}(x,x;\mu)=N_q\frac{(\alpha_q+\beta_q)^{(\alpha_q^{\alpha_q}+\beta_q^{\beta_q})}}{\alpha_q^{\alpha_q}\beta_q^{\beta_q}}
x^{\alpha_q}(1-x)^{\beta_q}f_{q/p}(x,\mu),
\label{eq:QS_function}
\end{align}
with the free parameters given in Table.~\uppercase\expandafter{\romannumeral1} in Ref.~\cite{Echevarria:2014xaa}.
As an approximation, the values of the strong coupling $\alpha_s$ are obtained at 2-loop order as
\begin{align}
\alpha_s(Q^2)&=\frac{12\pi}{(33-2n_f)\mathrm{ln}(Q^2/\Lambda^2_{QCD})}
\left\{{1-\frac{6(153-19n_f)}{(33-2n_f)^2}
\frac{\mathrm{ln}\mathrm{ln}(Q^2/\Lambda^2_{QCD})}{\mathrm{ln}(Q^2/\Lambda^2_{QCD})}}\right\}, \label{eq:alphas}
\end{align}
where $n_f=5$ and $\Lambda_{\mathrm{QCD}}=0.225\ \mathrm{GeV}$. We note that the running coupling in Eq.~(\ref{eq:alphas}) satisfies $\alpha_s(M_Z^2)=0.118$.

We still need to know the energy dependence of the Qiu-Sterman function $T_{q,F}(x,x;\mu)$ for calculating the spin-dependent differential cross section (\ref{eq:dcs_Sivers}).
We adopt two different approaches to deal with the scale dependence of $T_{q,F}$ for comparison. The first one (we label it as set 1) is to assume that $T_{q,F}$ is proportional to the usual unpolarized collinear PDF at any scale, that is, $T_{q,F}$ follows the DGLAP evolution as that of $f_1$, like the choice in Ref.~\cite{Echevarria:2014xaa}. The second one (we label it as set 2) is to adopt the parameterizations in Eq.~(\ref{eq:QS_function}) at the initial scale ($Q_0^2=2.4~\mathrm{GeV}^2$) and then evolve it to another scale $Q$ through QCD evolution using the evolution equation for $T_{q,F}$.
The evolution of $T_{q,F}$ has been studied extensively in literature~\cite{Kang:2012em,Kang:2008ey,Zhou:2008mz,Vogelsang:2009pj,Braun:2009mi,Ma:2011nd,
Schafer:2012ra,Ma:2012xn,Sun:2013hua,Zhou:2015lxa}.
In the second choice, for our purpose, we only consider the homogenous terms (the terms containing $T_{q,F}(x,x)$) in the evolution kernel as an approximation:
\begin{align}
P^{\mathrm{QS}}_{qq}\approx P^{f_1}_{qq}-\frac{N_c}{2}\frac{1+z^2}{1-z}-N_c\delta(1-z),
\label{eq:P_qq_Sivers}
\end{align}
where $P^{f_1}_{qq}$ being the evolution kernel of the unpolarized PDF
\begin{align}
P^{f_1}_{qq}=\frac{4}{3}\left(\frac{1+z^2}{(1-z)_+}+\frac{3}{2}\delta(1-z)\right).
\end{align}
In Ref.~\cite{Kang:2015msa}, similar choice (keeping the homogenous terms) was also adopted for the twist-3 fragmentation function $\hat H^{(3)}$ in the study of the Collins asymmetry.

\begin{figure}
\centering
\includegraphics[width=0.48\columnwidth]{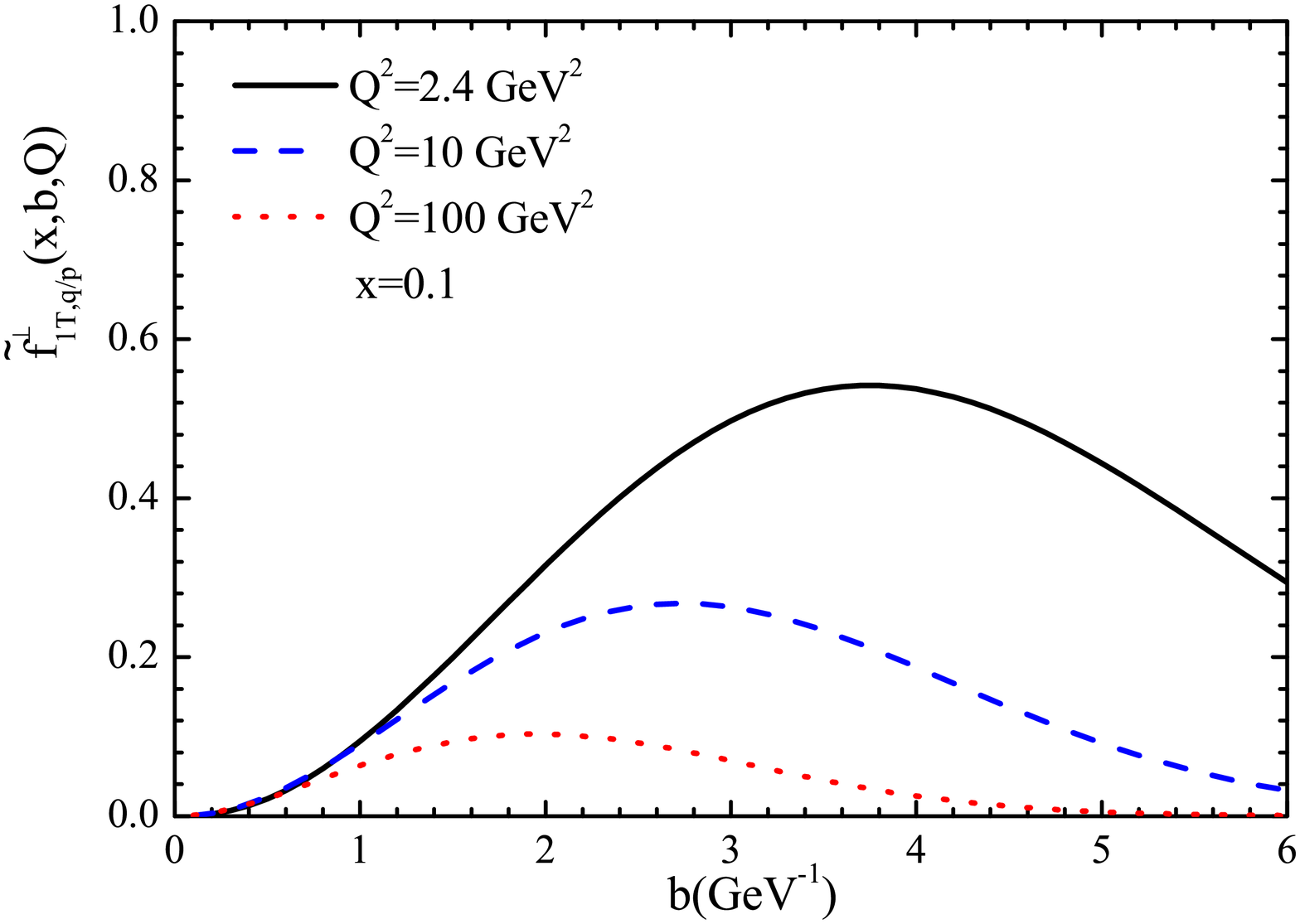}
\includegraphics[width=0.48\columnwidth]{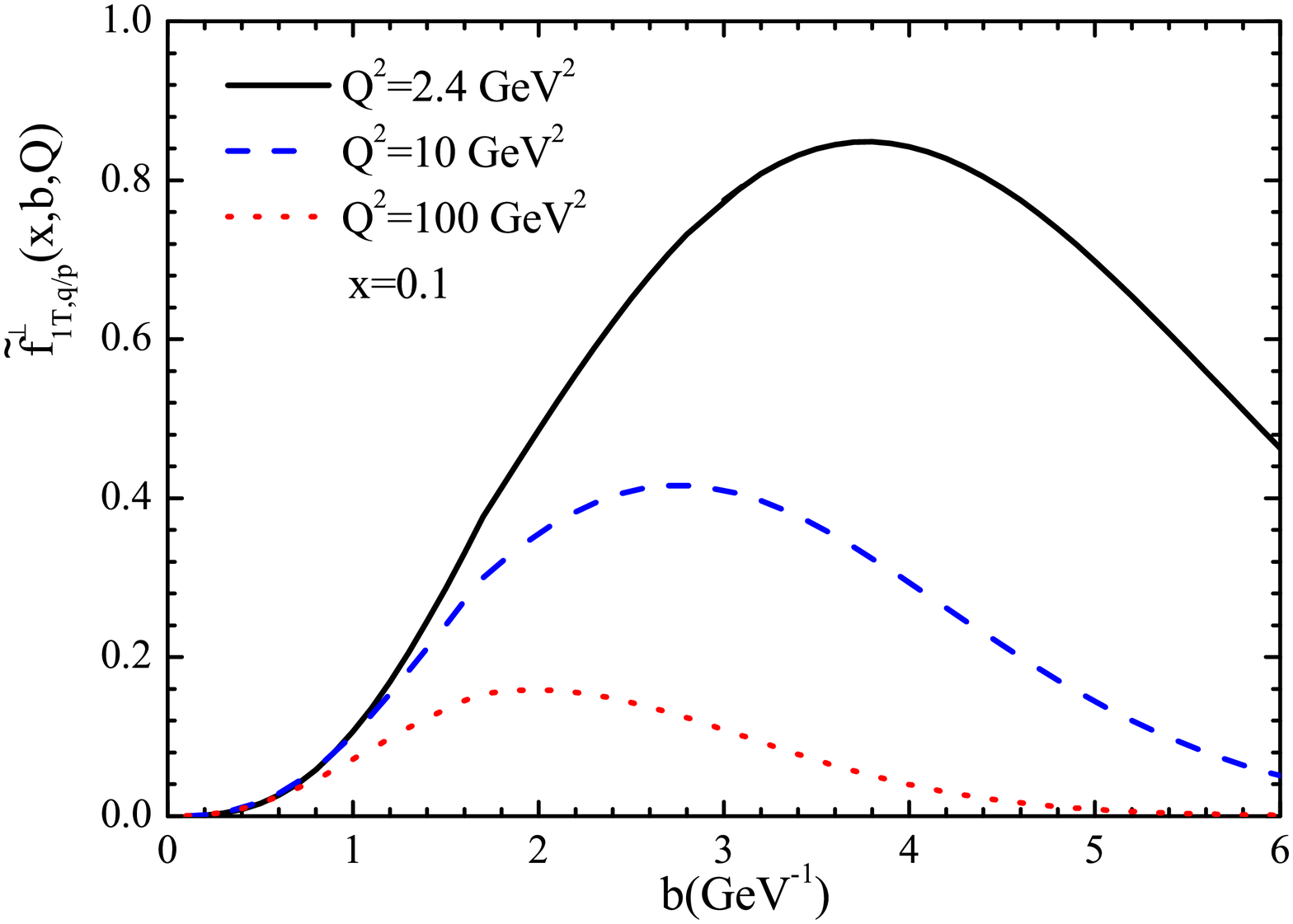}\\
\includegraphics[width=0.48\columnwidth]{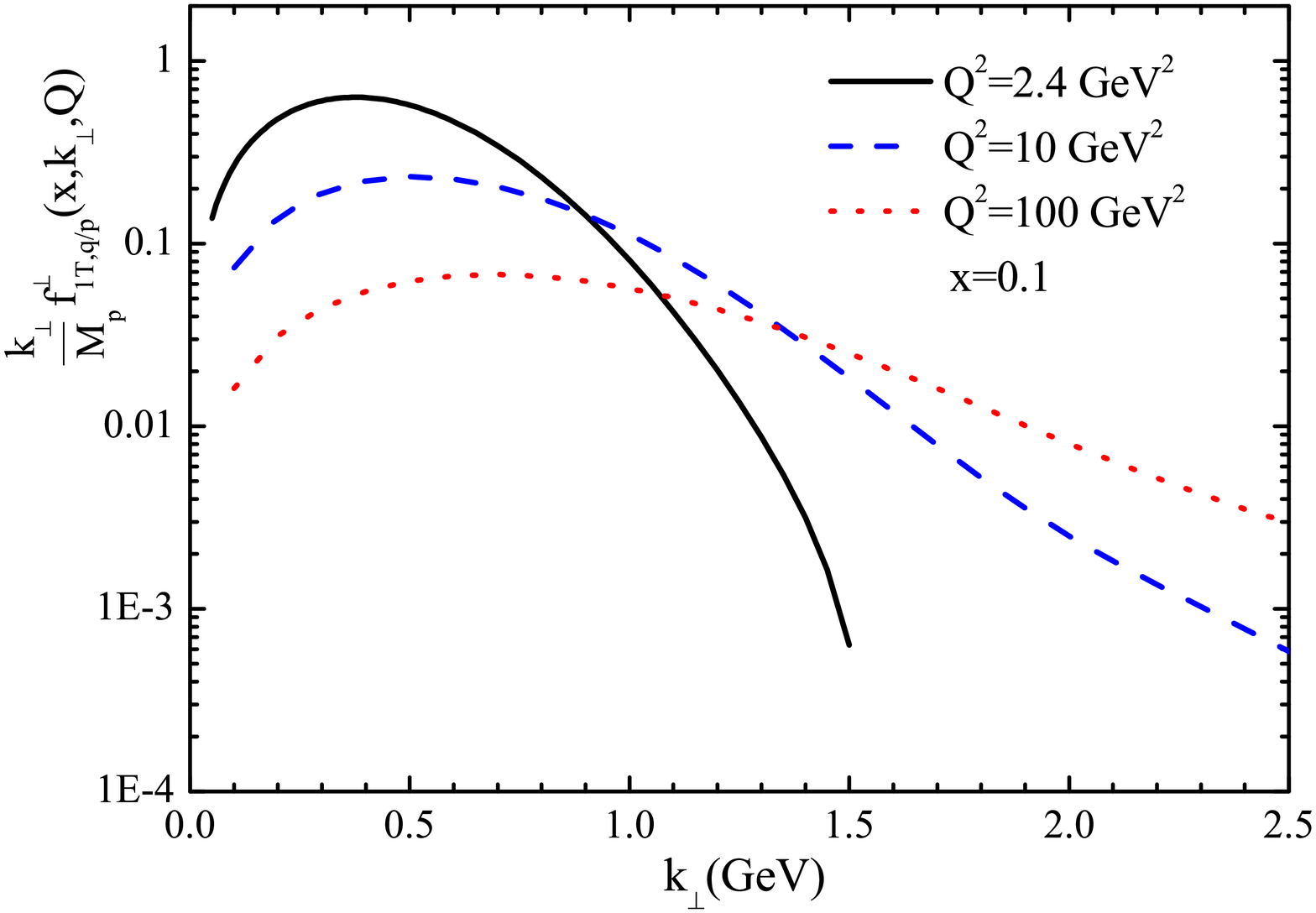}
\includegraphics[width=0.48\columnwidth]{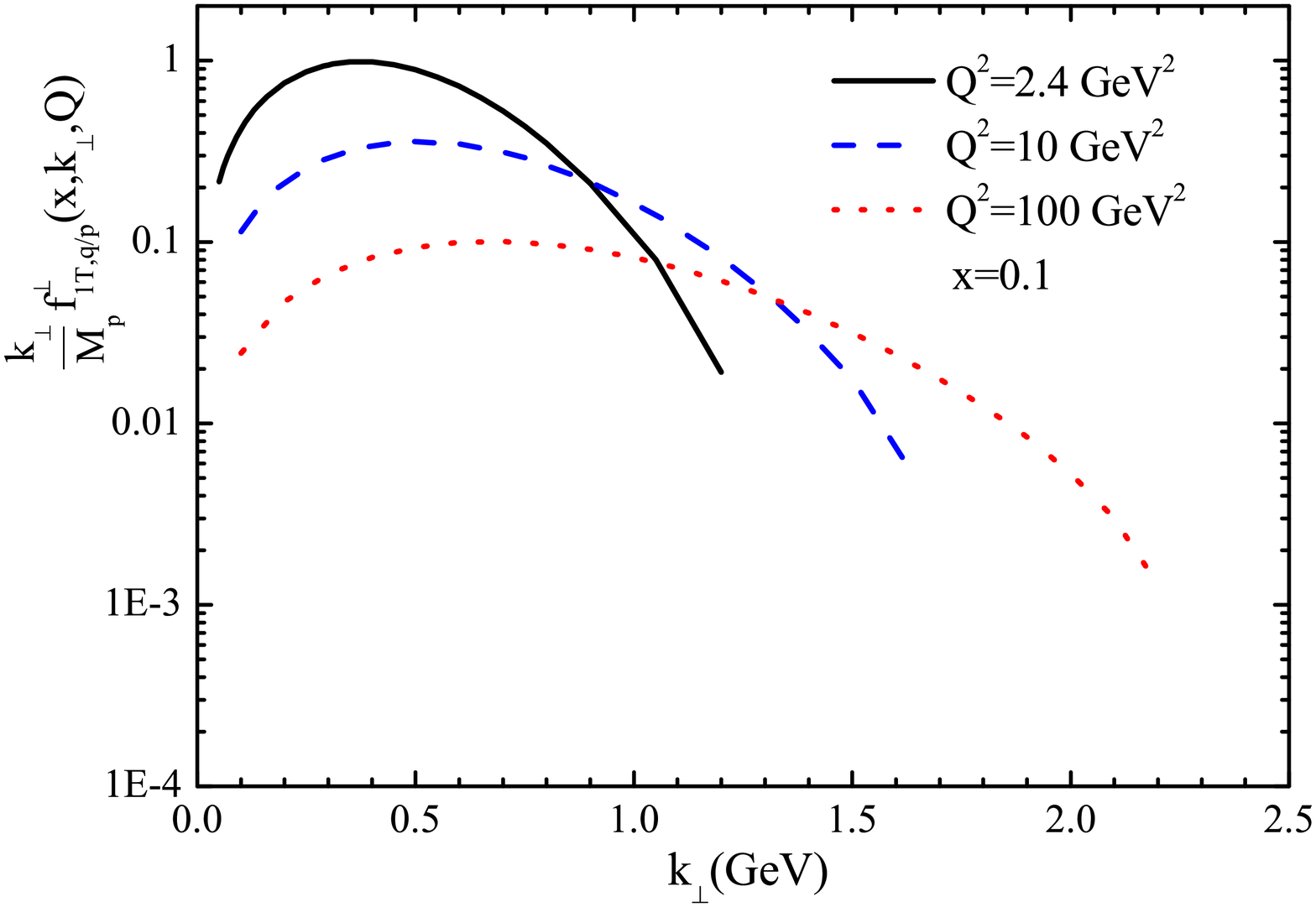}

\caption{Subtracted Sivers function for the up quarks in Drell-Yan in $b$-space (upper panels) and $k_\perp$-space (lower panels), at energies: $Q^2=2.4\ \mathrm{GeV}^2$~(solid lines), $Q^2=10\ \mathrm{GeV}^2$~(dashed lines) and $Q^2=100\ \mathrm{GeV}^2$~(dotted lines). The left and right panels plot the result of set 1 and set 2, respectively.}
\label{fig:Sivers_fun}
\end{figure}

To solve the QCD evolution numerically, we resort to the QCD evolution package HOPPET~\cite{Salam:2008qg,Salam:2008sz} and we custom the code to include the splitting function in Eq.~(\ref{eq:P_qq_Sivers}).
Using Eqs.~(\ref{eq:siv_b}) and (\ref{eq:QS_Q}), we apply the two choices for the scale dependence of $T_{q,F}$ to calculate the $b$-dependent and $k_\perp$-dependent Sivers function in Drell-Yan at different energy scales (we take the results of the $u$-quark Sivers function at fixed $x=0.1$ as an example): $Q^2=2.4 \mathrm{GeV}^2$, $Q^2=10 \mathrm{GeV}^2$ and $Q^2=100 \mathrm{GeV}^2$.

The numerical results are plotted in Fig.~\ref{fig:Sivers_fun}, in which the left panels show the results from the assumption that $T_{q,F}(x,x,\mu)$ is scaled as the collinear PDF $f_1(x,\mu)$, and the right panels depict the Sivers function calculated from the evolution kernel in Eq.~(\ref{eq:P_qq_Sivers}).
We find that the evolution effect is significant, i.e., it changes the shape and the size of the Sivers function at different $Q$ values. It also drives the peaks of the $b$-dependent curves to the lower $b$ region and the peak of $k_\perp$-dependent curves to higher $k_\perp$ region.
This indicates that the perturbative Sudakov form factor dominates in the low $b$ region at higher energy scales and the nonperturbative part of the TMD evolution becomes more important at lower energy scales.
Furthermore, generally the size of the Sivers function in set 2 is larger than that in set 1.
Besides, the $k_\perp$ tendency of the Sivers function in the two sets is different, namely, at larger $Q^2$ the sivers function in set 1 fall more slowly with increasing $k_\perp$ than that in Set 2.

\begin{figure}
\centering
\includegraphics[width=0.48\columnwidth]{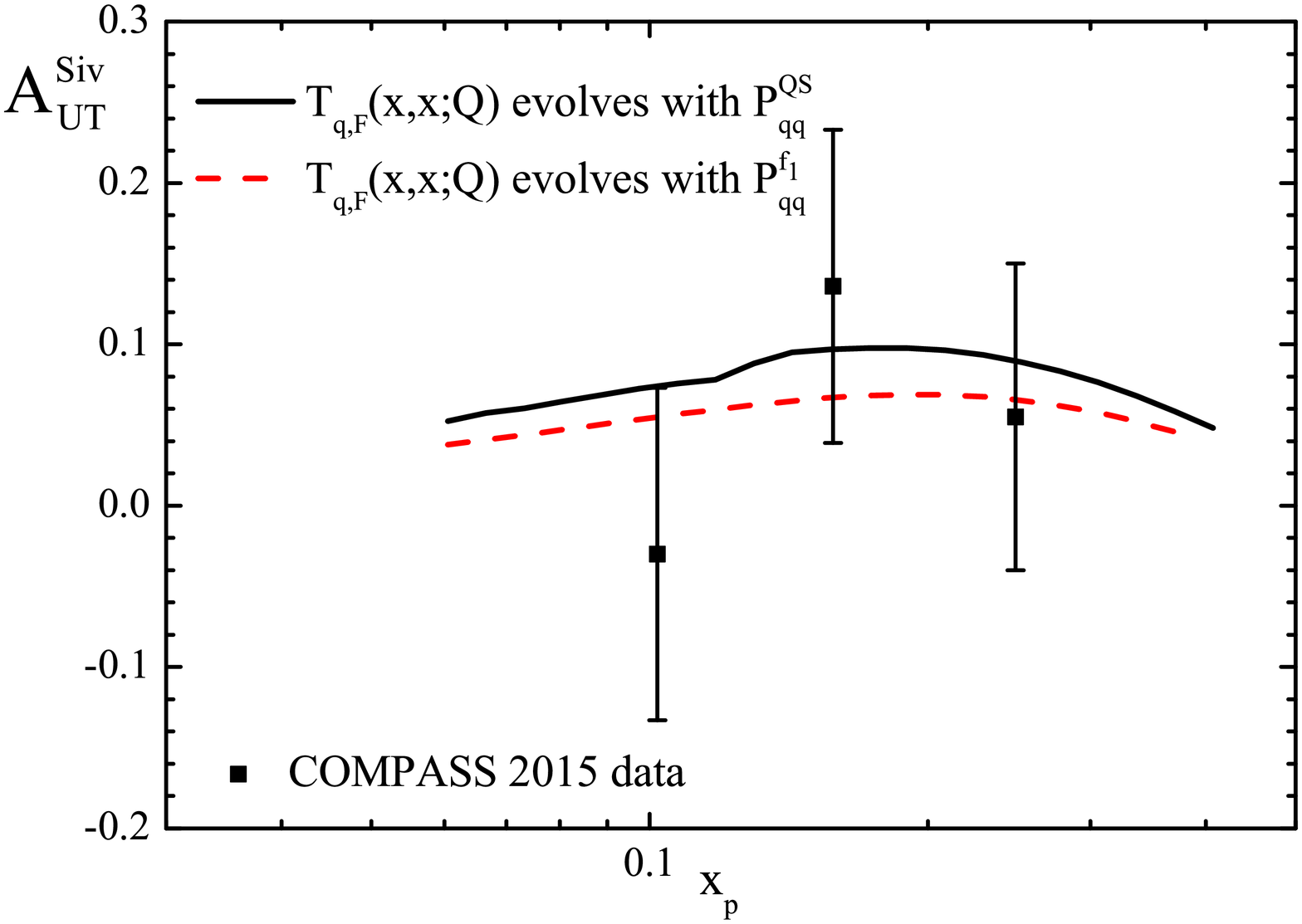}\quad
\includegraphics[width=0.48\columnwidth]{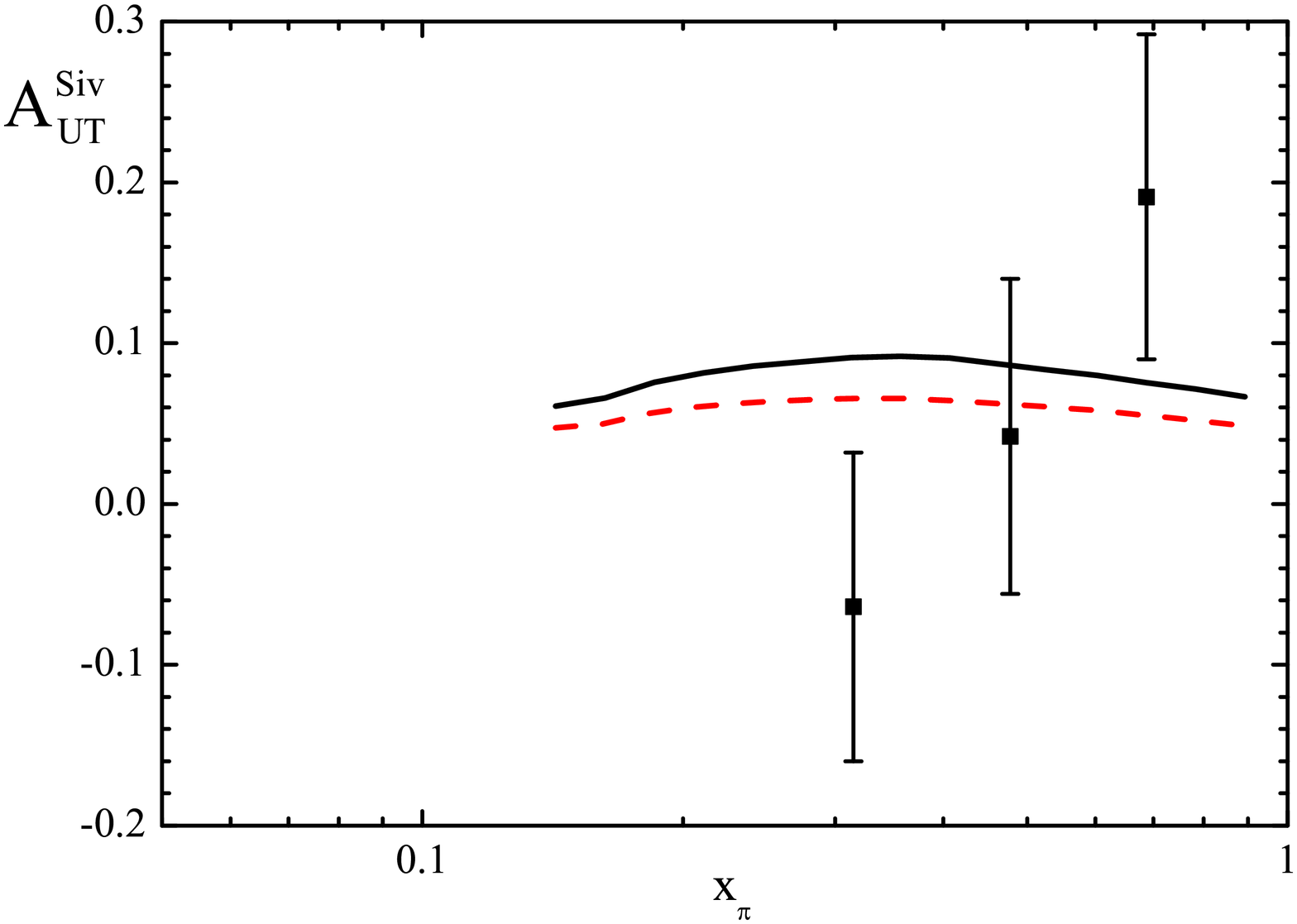}
\includegraphics[width=0.48\columnwidth]{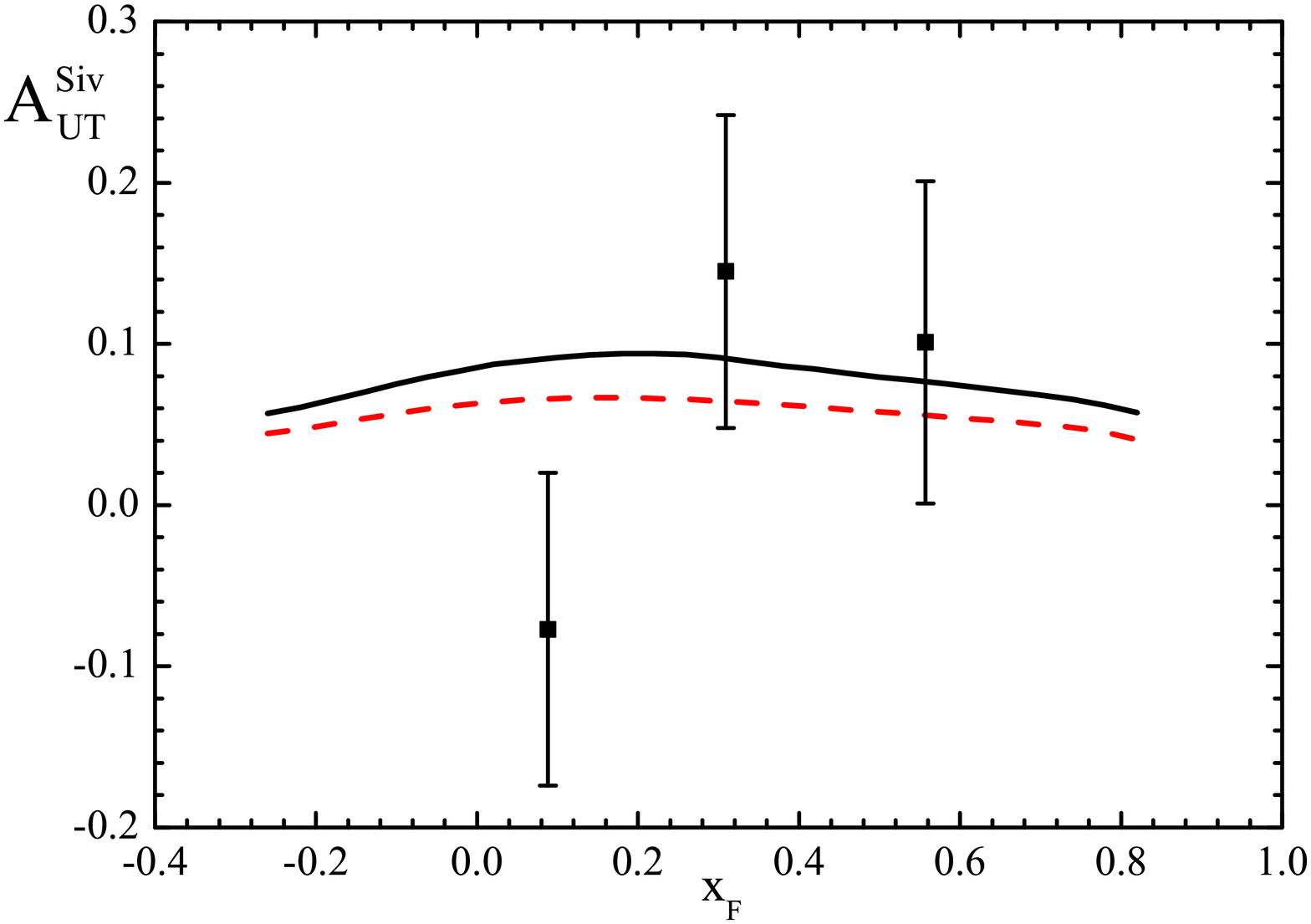}\quad
\includegraphics[width=0.48\columnwidth]{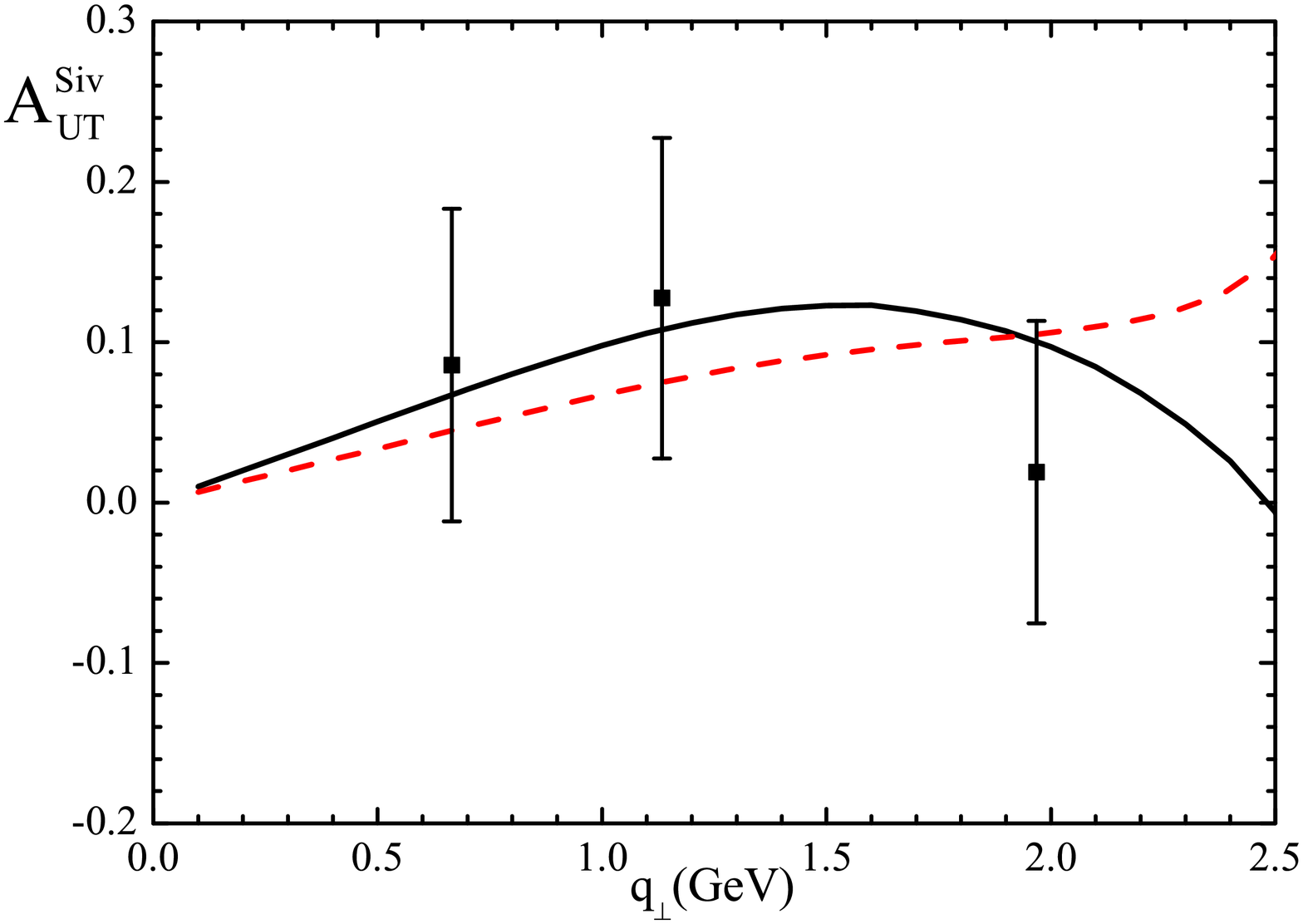}\\

\caption{The Sivers asymmetry for $\pi^-$ scattering off transversely polarized proton Drell-Yan process as functions of $x_p$~(upper left), $x_\pi$~(upper right), $x_F$~(lower left) and $q_\perp$~(lower right), compared with the COMPASS data~\cite{Aghasyan:2017jop}. The solid lines represent Sivers asymmetry with the Qiu-Sterman functions $T_{q,F}(x,x;Q)$ being proportional to the unpolarized PDF $f_1(x,Q)$. The dashed lines depict Sivers asymmetry considering Qiu-Sterman functions evolving through the splitting function in Eq.~(\ref{eq:P_qq_Sivers}).}
\label{fig:Sivers_asy}
\end{figure}

The COMPASS Collaboration at CERN has performed the first measurement of the Sivers asymmetry in the $\pi^- N$ Drell-Yan process~\cite{Aghasyan:2017jop}, using a $\pi^-$ beam with $P_\pi=\ 190\ \mathrm{GeV}$ colliding on a polarized $\mathrm{NH}_3$ target~\cite{Gautheron:2010wva,Aghasyan:2017jop}, which can serve as a transversely polarized nucleon target. The kinematical range covered at COMPASS in this measurement is as following
\begin{align}
&0.05<x_N<0.4,\quad 0.05<x_\pi<0.9, \quad
4.3\ \mathrm{GeV}<Q<8.5\ \mathrm{GeV},\quad s=357~\mathrm{GeV}^2,\quad -0.3<x_F<1.
\label{eq:cuts}
\end{align}

We apply Eqs.~(\ref{eq:asy_Sivers}), (\ref{eq:dcs_unp_id}) and (\ref{eq:dcs_Sivers}) to calculate the Sivers asymmetry $A_{UT}^{\textrm{Siv}}$ in the $\pi^- p$ Drell-Yan at the kinematics of COMPASS and plot the results in Fig.~\ref{fig:Sivers_asy}. To make the TMD factorization valid, the integration over the transverse momentum $q_\perp$ is performed in the region of $0<q_\perp<2~\mathrm{GeV}$, which is the same as the cut in Ref.~\cite{Sun:2013hua}. The upper panels of Fig.~\ref{fig:Sivers_asy} show the asymmetries as functions of $x_p$ (left panel) and $x_\pi$ (right panel); and the lower panels depict the $x_F$-dependent and $q_\perp$-dependent asymmetries, respectively. Among the plots, the dashed lines corresponds to the asymmetries obtained from the DGLAP evolution for $T_{q,F}(x,x;\mu)$, while the solid lines represent the results from the second choice of the evolution for $T_{q,F}(x,x;\mu)$, i.e, the one using the approximated kernel in Eq. (\ref{eq:P_qq_Sivers}) from the initial scale $Q_0^2= 2.4$ GeV$^2$. As a comparison, we also show the experimental data measured by the COMPASS Collaboration~\cite{Aghasyan:2017jop} with error bars in Fig.~\ref{fig:Sivers_asy}.

As shown in Fig.~\ref{fig:Sivers_asy}, in all the cases the Sivers asymmetry in the $\pi^- p$ Drell-Yan from our calculation is positive. The size of the asymmetry is around 0.05 to 0.10. This result is consistent with the COMPASS measurement shown in Fig.5 of Ref.~\cite{Aghasyan:2017jop} within the uncertainties of the asymmetry.
We also find that the asymmetry from the Sivers function in set 2 is more sizable than the one from set 1, and is more closer to the central values of the asymmetry measured by COMPASS. This is because that the Sivers function in set 2 is larger than the Sivers function in set 1.
Furthermore, compared to the asymmetry from set 1, the asymmetry from set 2 has a fall at larger $q_\perp$,  which is more compatible to the shape of $q_\perp$-dependent asymmetry of measured by the COMPASS Collaboration.
In conclusion, our analysis demonstrates that, combining the previous analysis of unpolarized pion TMD PDFs and that of the proton Sivers function within the TMD factorization and evolution, can lead to the Sivers asymmetry in $\pi^- N$ Drell-Yan which is consistent with the COMPASS measurement.

\section{Conclusion}
\label{Sec.conclusion}

In this work, we applied the formalism of the TMD factorization to study the Sivers asymmetry in the pion induced Drell-Yan process that is accessible at COMPASS.
We took into account the TMD evolution of the pion unpolarized distribution function as well as the proton Sivers function. For the former one, we carried out the nonperturbative Sudakov form factor of the pion TMD distributions extracted from the unpolarized $\pi N$ Drell-Yan data, while for the latter one, we adopted a parametrization of the nonperturbative Sudakov form factor that is universal and can describe the data of SIDIS, DY dilepton and W/Z boson production in $pp$ collisions.
We applied two different ways to treat the energy dependence of the Qiu-Sterman function which is proportional to first $k_\perp$ moment of the Sivers function. The first one is to assume the Qiu-Sterman function has the same scale dependence of the collinear unpolarized PDF, and the second one is to take the parametrization at the initial energy $Q_0$ and evolve it to another energy through an approximate evolution kernel for the Qiu-Sterman function containing the homogenous terms.
We then calculated the Sivers asymmetry in $\pi p$ Drell-Yan process at COMPASS as functions of the kinematical variables $x_p, x_\pi, x_F$ and $q_\perp$.
We find that the Sivers asymmetry calculated from the TMD evolution formalism is consistent with the COMPASS measurement.
Furthermore, different treatments on the scale dependence of the Qiu-Sterman function yield different sizes and shapes of the asymmetries.
Specifically, the calculation using an approximate evolution kernel for $T_{q,F}$ seems more compatible to the COMPASS data than the one using the DGLAP evolution.
Our study shows that, besides the TMD evolution effect, the scale dependence of the Qiu-Sterman function will play a role in the interpretation of the experimental data, and it should also be considered in the phenomenological studies.

\section*{Acknowledgements}
This work is partially supported by the NSFC (China) grant 11575043, by the Fundamental Research Funds for the Central Universities of China. X.~W. is supported by the Scientific Research Foundation of Graduate School of Southeast University (Grants No.~YBJJ1667).

\end{document}